\documentclass{article}
\usepackage{epsfig}

\def\real{{ {\hbox{$\displaystyle \kern-.2mm 1\kern-1mm \mbox{\rm R}\kern-.2mm$}}}}

\def\complex{{
    {\hbox{$\displaystyle\kern-.2mm {\rm C}\kern-1.5mm\raise.2mm
     \hbox{\vrule height6pt}\kern1.3mm$}}}}

\def\wPm{{$\widehat P$-matrix}}

\begin{document}

\title{Orbit Spaces in Superconductivity}

\author{Vittorino Talamini}

\date{INFN, Sez. di Padova, via Marzolo 8, 35131 Padova, Italy,
and\\
Dip. di Ingegneria Civile, Universit\`a di Udine, 33100 Udine, Italy\\
e-mail: vittorino.talamini@pd.infn.it\\}

\maketitle

\abstract{In the framework of Landau theory of phase transitions
one is interested to describe all the possible low symmetry
``superconducting'' phases allowed for a given superconductor
crystal and to determine the conditions under which this crystal
undergoes a phase transition. These problems are best described
and analyzed in the orbit space of the high symmetry group of the
``normal, non-superconducting'' phase of the crystal. In this
article it is worked out a simple example concerning
superconductivity, that shows the \wPm\ method to determine the
equations and inequalities defining the  orbit space and its
stratification. This approach is of general validity and can be
used in all physical problems that make use of invariant
functions, as long as the symmetry group is compact. \footnote{
Published in "International Workshop on Symmetry and Perturbation
Theory", Rome, 16-22/12/1998, eds. A. Degasperis and G. Gaeta,
World Scientific, Singapore (1999), 268-277.}}

\section{Introduction}
Landau theory of phase transitions~\cite{guf82,toltol,kosc} has
been sometimes used to study superconductor
crystals.~\cite{vol-gor,sgr-ueda} In that framework one deals with
a given potential function $V(x)$, defined in a finite dimensional
vector space $X$, that is invariant under transformations of a
compact group $G$ acting linearly on $X$: $V(gx)=V(x),\ \forall
g\in G,\ x\in X$. The vectors $x\in X$ are called {\em order
parameters}. All finite dimensional representations of a compact
group $G$ are unitary and completely reducible, and, given a
complex irreducible representation $\varphi$ of $G$, either
$\varphi$ itself, or the direct sum $\varphi+\bar{\varphi}$, where
$\bar{\varphi}$ is the complex conjugate representation of
$\varphi$, is equivalent to a real orthogonal irreducible (on the
reals) representation of $G$.~\cite{naim,hamer} In Landau theory,
in a first approximation, sufficient to determine all possible
symmetry phases allowed for the system, one considers $X$ hosting
a representation of $G$ equivalent to a real orthogonal
irreducible representation.~\cite{toltol} This implies that no
vector $x\neq 0$ is left invariant by all elements of $G$, so the
action is {\em effective}. In all generality, one may then assume
that $X\equiv\real^n$ and that the group $G$ is a real orthogonal
subgroup of O($n,\real$). In the following the irreducibility
condition will never be used, but we still require effective
orthogonal actions in $\real^n$.\\ In this Introduction some of
the mathematical tools developed in~\cite{as2,st-cmp} are
reviewed; they are used in Section 2 to determine the symmetry
phases allowed for the physical system considered. The method
exposed modifies slightly that one proposed by Gufan.~\cite{guf71}
The physical results here obtained are not new,~\cite{guf94} but I
would instead point the attention of the reader to the
mathematical method used to derive the orbit space stratification
and the list of phase transitions allowed for a given potential.

 If $x_0$ is the minimum point of $V(x)$, then the
system is in equilibrium in a state compatible with the order
parameters $x_0$. If $x_0\neq 0$ then the symmetry of the system
is lower than $G$, it is reduced to $G_{x_0}=\{g\in G \mid g
x_0=x_0\}$, the {\em isotropy subgroup} of $x_0$, realizing a {\em
spontaneous symmetry breaking}.\\ Moreover, if $x_0\neq 0$, the
{\em orbit} $\Omega(x_0)=\{g x_0, \ g\in G\}$ contains points
$x\neq x_0$ where $V(x)=V(x_0)$, so the low symmetry equilibrium
states are always degenerate. The following relation
holds:~\cite{bred,kosc} $G_{g x}=g  G_{x} g^{-1},\ \forall g\in
G,\ x\in \real^n$, then all points $x' \in \Omega(x)$ have
isotropy subgroups in the same conjugacy class $[G_x]$ and
viceversa, all groups $ G' \in [G_x]$ are isotropy subgroups of
some points $x' \in \Omega(x)$. The conjugacy class $[G_{x}]$ is
called the {\em (isotropy) type} of $\Omega(x)$. Two conjugated
isotropy subgroups are physically equivalent because they can be
thought to be the same group seen from two rotated systems of
reference, of course we are here considering $G$ orthogonal. Given
an isotropy subgroup $H \subseteq G$, the set of points
$\Sigma_{[H]}=\{x\in \real^n \mid G_x \in [H]\}$ is called the
{\em stratum} of type $[H]$. The strata are disjoint and in a one
to one correspondence with the {\em symmetry phases} that may in
principle be accessible to the system. Orbits and strata can be
partially ordered according to their types: the type $[H]$ is said
to be {\em smaller} than the type $[K]$, $[H]<[K]$, if $H' \subset
K'$ for some $H'\in [H]$ and $K'\in [K]$. If $G$ is compact and is
acting orthogonally in $\real^n$, the number of strata is
finite,~\cite{mostow} p. 444, and there is a unique stratum of the
smallest type,~\cite{bred} called the {\em principal stratum}.

 The values assumed by the potential $V$ generally
depend, in addition to $x$, on other variables, like temperature,
pressure, atomic concentrations, etc., and the absolute minimum of
$V$ falls in a point $x_0$ that depends on these variables.
Sometimes, when these variables change their values, $x_0$ may
change stratum, and in this case the system undergoes a {\em phase
transition}.

The {\em orbit space} of $G$ is the quotient space $\real^n/G$,
whose points represent whole orbits of $\real^n$. A stratum of the
orbit space is formed by all the points that represent orbits in a
same stratum of $\real^n$. The invariant functions, e.g. $V(x)$,
are naturally defined in the orbit space because they are constant
on the orbits, so, to eliminate the degeneracy associated to the
symmetry, one is lead to study the invariant functions in the
orbit space.

To determine concretely the equations and inequalities that define
the orbit space and its strata one may take advantage of the \wPm\
and its properties.~\cite{as1,as2} By Hilbert's
theorem,~\cite{hilb,weyl} the algebra $\real[\real^n]^G$ of real
$G$-invariant polynomials defined in $\real^n$ has a finite number
$q$ of generators: $p_1,\ldots,p_q \in \real[\real^n]^G$, forming
an {\em integrity basis} for $\real[\real^n]^G$. This means that
$$ f(x)=\widehat f(p_1(x),\ldots,p_q(x))\qquad \forall\,f\in
\real[\real^n]^G,\quad x\in\real^n, $$ with $\widehat f$ a real
polynomial in $q$ variables. Actually, the preceding formula holds
true for any $G$-invariant $C^\infty$ function $f(x)$, then
$\widehat f$ is a $C^\infty$ function of $q$
variables.~\cite{bierst-jdg,schw-top} If no subset of
$\{p_1,\ldots,p_q \}$ is still an integrity basis for
$\real[\real^n]^G,$ then the set $\{p_1,\ldots,p_q \}$ forms a
{\em minimal integrity basis} (MIB) for $\real[\real^n]^G$. There
is some arbitrariness in the choice of the MIB, however all its
elements can be chosen to be homogeneous and one may order the
$p_a,\ a=1,\ldots,q,$ according to their degrees $d_a=\deg
p_a(x)$, for example $d_a\geq d_{a+1}$. If $G$ acts effectively in
$\real^n$, then $d_q= 2$, and, for the orthogonality of $G$, one
may take $\ p_q(x)=|x|^2=\sum_{i=1}^n x_i^2$.

The vector map $p:\real^n \to \real^q:x \to p(x)=(p_1(x),
p_2(x),\ldots,p_q(x))$, called the {\em orbit map}, is constant in
each orbit $\Omega\subset \real^n$ because the $p_a(x)$ are
$G$-invariant functions. The point $p=p(x)\in\real^q$ is the image
in $\real^q$ of the orbit $\Omega(x)$ of $\real^n$. No other orbit
$\Omega'\neq \Omega$ of $\real^n$ is mapped to the same point
$p\in\real^q$ because the MIB separates the
orbits.~\cite{schw-top,as2} The image of $\real^n$ through the
orbit map is the set $${\cal S}=p(\real^n) \subset\real^q$$ that
can be identified with the orbit space of the
$G$-action.~\cite{schw-top}\\ The strata $\sigma$ of ${\cal S}$
are the images of the strata $\Sigma$ of $\real^n$ through the
orbit map: $\sigma=p(\Sigma)$. The principal stratum $\sigma_p$ of
${\cal S}$  is connected, open and dense in ${\cal
S}$.~\cite{mont-yang} Given two strata $\Sigma$ and $\Sigma'$ of
$\real^n$, if $\Sigma'$  is of greater type than $\Sigma$, then
$\sigma'=p(\Sigma')$ lie in the boundary of
$\sigma=p(\Sigma)$.~\cite{bierst-top} The smallest stratum of
${\cal S}$ is $\sigma_{[G]}$, image of the origin of $\real^n$ and
is located at the origin of $\real^q$. Given any plane $\Pi_r$ of
$\real^q$ with equation $p_q=r^2>0$, ${\cal S}\cap \Pi_r$, is a
{\em non-empty compact connected} section of ${\cal S}$ that
contains all strata of ${\cal S}$ except
$\sigma_{[G]}$.~\cite{as2,st-cmp} This section is sufficient to
describe the whole shape of ${\cal S}$, because moving $\Pi_r$
toward the infinite or toward the origin of $\real^q$, by changing
$r^2$, ${\cal S}\cap \Pi_r$ expands or contracts, but maintains
its topological shape and stratification.

Let's define the $q \times q$ symmetric and positive semi-definite
matrix $P(x)$, with elements $$   P_{ab}(x) = \nabla p_a(x)\cdot
\nabla p_b(x)=\sum_{i=1}^n \frac{\partial p_a(x)}{\partial
x_i}\frac{\partial p_b(x)}{\partial x_i}\quad
 a,b=1,\ldots,q.$$
The matrix elements $P_{ab}(x)$ are real homogeneous polynomial
functions of $x$ with $\deg P_{ab}(x)=d_a+d_b-2$, and
$P_{qa}(x)=P_{aq}(x)=2d_a p_a(x)$. A less immediate property is
the $G$-invariance; it follows from the orthogonality of $G$ and
from the covariance of the gradients of $G$-invariant functions:
$$P_{ab}(g x)=\nabla p_a(g x)\cdot\nabla p_b(g x)= g \nabla
p_a(x)\cdot g \nabla p_b(x)= \nabla p_a(x)\cdot \nabla
p_b(x)=P_{ab}(x).$$ Then, $P_{ab}(x)$ can be expressed in terms of
the  MIB: $$ P_{ab}(x)=\widehat P_{ab}(p_1(x),\ldots,p_q(x)),\quad
\forall x \in \real^n,\quad a,b=1,\ldots,q.$$ The matrix $\widehat
P(p)$, defined in $\real^q$, is called the {\em \wPm}. When  $p=
p(x)$, then $\widehat P(p)$ coincides with $P(x)$.

Let's call a polynomial $\widehat f(p)$ {\em $w$-homogeneous} of
{\em weight} $d$ if the polynomial  $f(x)=\widehat f(p(x))$ is
homogeneous of degree $d$, and define the {\em surface of the
relations}: ${\cal Z}=\{p\in\real^q\mid \widehat f_i(p)=0,\
i=1,\ldots,k\}$, where $\widehat f_i(p_1(x),\ldots,p_q(x))= 0,\
i=1,\ldots,k,$ are $k$ algebraically independent homogeneous
polynomial relations, called {\em syzygies}, existing among the
$p_a(x)$. If there are no syzygies, then $k=0$ and ${\cal Z}
\equiv \real^q$.

The main properties of a \wPm\ are the following:~\cite{as2,ps}
\begin{itemize}
\item $\widehat P(p)$ is a real, symmetric $q \times q$ matrix.
The matrix elements $\widehat P_{ab}(p)$ are $w$-homogeneous
polynomials of the $p_a$ of weight $d_a+d_b-2$, and $\widehat
P_{qa}(p) =\widehat P_{aq}(p) = 2 d_a p_a,\ \forall a=1,\ldots,q.$
\item  $\widehat P(p)$ is positive semi-definite in ${\cal S}$.
Precisely, ${\cal S}=\{p\in {\cal Z}\mid\widehat P(p)\geq 0\}$;
\item
Given a stratum $\sigma\subset {\cal S}$, and $p\in\sigma$,
$\dim\sigma=\mbox{rank}\widehat P(p)$.
\end{itemize}
The \wPm\ completely determines the orbit space ${\cal S}$ and its
stratification. Defining ${\cal S}_d$ as the union of all
$d$-dimensional strata of ${\cal S}$, one has: $${\cal S}_d=\{p\in
{\cal Z} \mid \widehat P(p)\geq 0,\ \mbox{rank}\widehat
P(p)=d\},$$  so, ${\cal S}=\overline{{\cal S}_{q-k}}$, where
$q-k=\dim{\cal Z}$ and $k$ is the number of syzygies. As all these
positivity and rank conditions are expressed through a finite
number of polynomial equations and inequalities, ${\cal S}$ is a
semi-algebraic set with a finite number of strata (confirming
Mostow result~\cite{mostow}).

The polynomials defining the strata $\sigma$ of ${\cal S}$ of
dimension $q-1$ (that exist only if there is at most one syzygy)
must satisfy a set of differential relations that characterize
them.~\cite{sar-mpl,st-cmp} Let $a(p)$ be an irreducible
polynomial defining such a stratum of ${\cal S}$. Then the
following {\em master} relations hold true:~\cite{sar-mpl}
$$\sum_{b=1}^q {\widehat P}_{ab}(p) \frac{\partial a(p)}{\partial
p_b}= \lambda_{a}(p) a(p) \qquad a=1,\ldots,q,$$ where the
$\lambda_{a}(p)$ are $w$-homogeneous polynomials of weight
$d_a-2$. The polynomials $a(p)$ that satisfy the above master
relations are called {\em active}. They are always factors of
$\det \widehat P(p)$.~\cite{st-cmp} The $(q-1)$-dimensional
stratum $\sigma$ is located in the interior of
$\overline{\sigma}=\{p\in \real^q \mid a(p)=0, \widehat P(p)\geq 0
\}$. The border $\overline{\sigma}\backslash \sigma$ of $\sigma$
is the union of lower dimensional strata whose defining equations
and inequalities can be found simplifying the system $\{
a_1(p)=a_2(p)=0, \widehat P(p)\geq 0 \}$, where $a_1(p)$ and
$a_2(p)$ are two irreducible active polynomials.

As the $\widehat P$-matrices completely define the orbit spaces
and their stratification, the classification of $\widehat
P$-matrices implies the classification of orbit spaces. All
$\widehat P$-matrices of dimension $q\leq 4$ corresponding to {\em
coregular} groups (i.e. with no syzygies) have been determined and
classified.~\cite{st-cmp,st-jgtp} This classification is obtained
independently from the specific group structures by making use
only of some general properties of orbit spaces. It is found that
many different groups, no matter if they are finite groups or
compact Lie groups, have the same \wPm\ and the same orbit
space.~\cite{st-jmp}

\section{A concrete example analyzed in the orbit space}

The example here proposed was first studied by Gufan~\cite{guf94}
and concerns the study of the possible symmetry phases of electron
pairs with non-zero total orbital momentum in hexagonal
superconducting crystals, in the approximation of strong
crystalline field and strong spin-orbit interaction.\\
In~\cite{guf94} the order parameters are chosen to be two complex
numbers, $\eta_{+}$ and $\eta_{-}$, and the symmetry group is
$\widehat Y=\widehat G\otimes \widehat R \otimes \widehat
U_1(\alpha)$, where $\widehat G$ is the point symmetry group of
the crystal, $\widehat R$ is the time reversal group and
$U_1(\alpha)$ is the rotation group of the complex plane. Here we
consider only the case $\widehat G=C_{3v}$. The group $\widehat Y$
acts in the four dimensional complex space of vectors of the
following form: $$\eta= \left(\begin{array}{c} \eta_{+}\\
\eta_{-}\\ \eta_{-}^\ast\\ \eta_{+}^\ast\\
\end{array} \right), $$ where $^\ast$ denotes complex conjugation, and the
group generators have the following matrix expressions: $$
\begin{array}{cccc}
C_3^{1z} & \sigma^v & U_1(\alpha)  & R \\
\left(\begin{array}{c}
e^{2 \pi i/3}\\
e^{-2 \pi i/3}\\
e^{2 \pi i/3}\\
e^{-2 \pi i/3}\\
\end{array}
\right) &
\left(\begin{array}{cccc}
0&1&0&0\\
1&0&0&0\\
0&0&0&1\\
0&0&1&0\\
\end{array}
\right) & \left(\begin{array}{c}
e^{i\alpha}\\
e^{i\alpha}\\
e^{-i\alpha}\\
e^{-i\alpha}\\
\end{array}
\right) &
\left(\begin{array}{cccc}
0&0&1&0\\
0&0&0&1\\
1&0&0&0\\
0&1&0&0\\
\end{array}
\right), \\
\end{array}
$$ where   $C_3^{1z}$ and $U_1(\alpha)$ are diagonal matrices, of
which only the diagonal elements are shown.\\ The MIB for this
group has 3 elements that can be written in the following real
form: $$J_3=|\eta_{+}|^2+|\eta_{-}|^2,\quad J_2=|\eta_{+}|^2
|\eta_{-}|^2,\quad J_1=\eta_{+}^3
{\eta_{-}^\ast}^3+{\eta_{+}^\ast}^3 \eta_{-}^3.$$ Of course one
may consider the action of $\widehat Y$ in $\complex^2$, but this
action would not be linear, the invariants would not be polynomial
functions of $\eta_{+}$ and $\eta_{-}$, and the theory exposed in
the Introduction would not be valid. \\ When a unitary
representation of a compact group is equivalent to a real
orthogonal representation, one may always choose the MIB's for the
two cases (unitary and orthogonal) to be formed by the same real
polynomial functions.~\cite{st-jmp} This means that the orbit map
defines in the two cases the same subset ${\cal S}\subset
\real^q$, that is the same orbit space.

\begin{figure}[h]
\begin{center}
\includegraphics{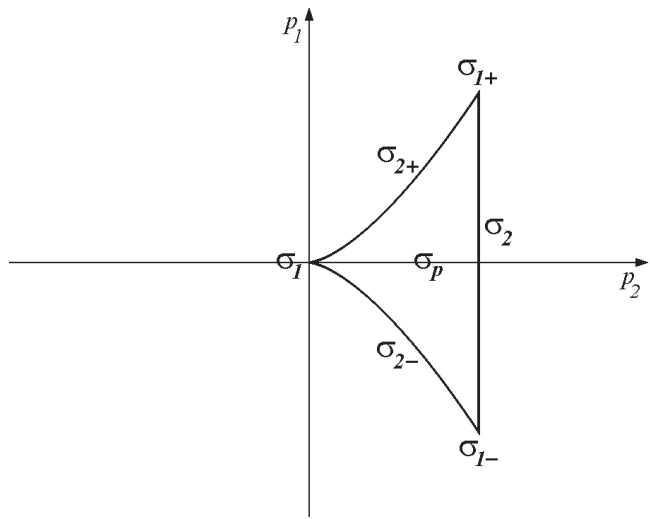}
 \caption{\protect\footnotesize
Section of the orbit space.\label{fig}}
\end{center}
\end{figure}

To adapt this scheme to that one reviewed in the Introduction, one
has to consider an orthogonal action of $\widehat Y$ in the real
space $\real^4$. Introducing 4 real variables $x_{1,2}$ and
$y_{1,2}$, one may put $\eta_{+}=x_1+i y_1$ and $\eta_{-}=x_2+i
y_2$, and consider a vector $x\in\real^4$ of the following form:
$$x= \left(\begin{array}{c} x_1\\ y_1\\ x_2\\ y_2\\
\end{array}
\right). $$ This defines a one-to-one correspondence between the
complex space of the $\eta$ and $\real^4$. The group $\widehat Y$
acts in $\real^4$ with the following matrix generators: $$
\begin{array}{cc}
C_3^{1z} & \sigma^v \\
\left(\begin{array}{cccc} -\frac12&-\frac{\sqrt{3}}{2}&0&0\\
\frac{\sqrt{3}}{2}&-\frac12&0&0\\
0&0&-\frac12&\frac{\sqrt{3}}{2}\\
0&0&-\frac{\sqrt{3}}{2}&-\frac12\\
\end{array}
\right) &
\left(\begin{array}{cccc} 0&0&1&0\\ 0&0&0&1\\ 1&0&0&0\\
0&1&0&0\\
\end{array}
\right)
\end{array}$$
$$
\begin{array}{cc}
U_1(\alpha)  & R \\

\left(\begin{array}{cccc} c&-s&0&0\\ s&c&0&0\\
0&0&c&-s\\ 0&0&s&c\\
\end{array}
\right) & \left(\begin{array}{cccc} 0&0&1&0\\ 0&0&0&-1\\ 1&0&0&0\\
0&-1&0&0\\
\end{array}
\right), \\
\end{array}
$$
where $c=\cos\alpha$ and $s=\sin\alpha$.\\ The MIB can be easily
written down in terms of the real variables:
\begin{eqnarray*}
p_3(x)&=&x_1^2+x_2^2+y_1^2+y_2^2,\\
p_2(x)&=&4(x_1^2+y_1^2)(x_2^2+y_2^2),\\ p_1(x)&=&16(x_1 x_2+y_1
y_2)(x_1^2 x_2^2-3 x_1^2 y_2^2+8 x_1 x_2 y_1 y_2-3 x_2^2
y_1^2+y_1^2 y_2^2),
\end{eqnarray*}
where the numeric factors have been introduced for later
convenience.\\ To find out the \wPm\ one has to calculate all
gradients of the $p_a(x)$, make their scalar products, and express
these products in terms of the $p_a(x)$. One finds, at the end,
the following \wPm: $$\widehat P(p)=\left(\begin{array}{ccc} 144\,
p_2^2\, p_3\ & 24\, p_1\, p_3\ & 12\, p_1\\ 24 \,p_1\, p_3\ &16
\,p_2\, p_3\  & 8 \, p_2\\ 12 \,p_1\ & 8  \,p_2\     & 4 \, p_3\\
\end{array}
\right), $$ whose determinant is $$\det \widehat P(p)= 2304\,
p_3\, (p_1^2-4 p_2^3)(p_2-p_3^2).$$ The $p_a(x)$ are algebraically
independent, so the principal stratum is the subset of $\real^3$
where $\widehat P(p)> 0$, and all other strata are contained in
the surface of equation $A(p)=0$, where $A(p)$ is the product of
all the irreducible active factors of $\det \widehat P(p)$. Using
the master relations, one finds: $$A(p)=(p_1^2-4
p_2^3)(p_2-p_3^2). $$ The section of the surface $A(p)=0$ in a
plane $p_3=$constant is plotted in Figure \ref{fig}. Its interior
hosts the section of the principal strata $\sigma_p$ and its
border contains the sections of all other strata, except the
origin.
The complete list of the strata of ${\cal S}$ is reported in Table
\ref{tab1}, where the defining equations and inequalities of the
bordering strata are found by examining $\mbox{rank} \widehat
P(p)$ in the surface $A(p)=0$.
\begin{table}[h]
\begin{center}
\caption{Strata of the orbit space \label{tab1}}
\begin{tabular}{|cc|}
\hline Stratum         & Defining relations         \\ \hline
$\sigma_0$     & $p_1 = p_2 =p_3 =0$
\\ $\sigma_1      $ & $p_1=p_2=0,\ p_3>0$    \\ $\sigma_{1+}$  &
$p_1=2p_3^3,\ p_2=p_3^2,\ p_3>0$  \\ $\sigma_{1-} $ &
$p_1=-2p_3^3,\ p_2=p_3^2,\ p_3>0$         \\ $\sigma_2 $    &
$|p_1|<2p_3^3,\ p_2=p_3^2,\ p_3>0$        \\ $\sigma_{2+}$  &
$p_1=2p_2^{\frac32},\ 0<p_2<p_3^2,\ p_3>0$         \\
$\sigma_{2-}$  & $p_1=-2p_2^{\frac32},\ 0<p_2<p_3^2,\ p_3>0$ \\
$\sigma_p$     & $p_1^2<4p_2^3,\ 0<p_2<p_3^2,\ p_3>0$  \\ \hline
\end{tabular}
\end{center}
\end{table}

\noindent Table \ref{tab2} lists for each stratum $\sigma$, some
conveniently chosen points $\overline{x}$ and the generators of
their isotropy subgroups $G_{\overline{x}}$. One may note that the
points in the border of a stratum $\sigma$ have isotropy subgroups
containing those of points of $\sigma$, modulo a conjugation. Just
to see this fact, in two cases are reported two different points
in a same orbit, related by the group element $U_1(\frac{
\pi}{2})$.
\begin{table}[h]
\begin{center}
\caption{Isotropy types of the strata\label{tab2}}
\begin{tabular}{|ccc|}
\hline Stratum         & Typical point $\overline{x}$    &
Generators of the isotropy subgroup of $\overline{x}$      \\
\hline $\sigma_0$     &  (0,0,0,0)            &
$U_1(\alpha);\;C_3^{1z};\;\sigma^v;\;R $      \\ $\sigma_1
$&(0,1,0,0)            & $U_1(-\frac23\pi)C_3^{1z};
\;U_1(\pi)\sigma^v R  $\\ $\sigma_{1+}$ & (0,1,0,1)
&$\sigma^v;U_1(\pi)R   $       \\ $\sigma_{1-} $ & (-1,0,1,0) &  $
U_1(\pi)\sigma^v;\;U_1(\pi)R    $ \\
           &  (0,1,0,-1)            &  $  U_1(\pi)\sigma^v;\;  R $ \\
$\sigma_2 $    &  (-1,1,1,1)            &     $U_1(\pi)R $   \\
           &  (1,1,1,-1)            &     $R $   \\
$\sigma_{2+}$  &  $(0,1,0,2)$          &    $U_1(\pi)\sigma^v R$
\\ $\sigma_{2-}$  &  $(0,1,0,-2)$          & $U_1(\pi)\sigma^v R$
\\ $\sigma_p$     &  $(-1,2,1,1)$ & $E$           \\ \hline
\end{tabular}
\end{center}
\end{table}

Once the orbit space is described, one may list all second order
phase transitions that may in principle be allowed for the system,
a second order phase transition may in fact take place only
between bordering strata because it is a consequence of a
continuous variation of the variables that appear in the
potential.~\cite{guf82,toltol,kosc} In the example here
considered, the possible second order phase transitions are the
following ones: $$\begin{array}{ccccccc}
 \sigma_0,\sigma_p    &\longleftrightarrow & \mbox{all other strata}& \qquad
 \sigma_{2+} &\longleftrightarrow & \sigma_1,\,\sigma_{1+}\\
 \sigma_2    &\longleftrightarrow & \sigma_{1+},\,\sigma_{1-}& \qquad
 \sigma_{2-} &\longleftrightarrow & \sigma_{1},\,\sigma_{1-}\\
\end{array}$$
The conditions under which the phase transitions are realized
depend on the form assumed for the potential when this is written
in terms of the MIB and not all the transitions listed above do
actually take place when a given form of the potential is
specified. It is not possible to examine here in detail the phase
transitions scenario associated to a given potential $\widehat
V(p)$ and I shall here only remind the basic points of this
analysis.~\cite{st-jmp,guf94} One first has to write down
$\widehat V(p)$ in terms of the MIB. Generally, one uses a
polynomial approximation for $\widehat V(p)$: $\widehat V(p)=a_1
p_3+a_2 p_3^2+\ldots+b_1 p_2+\ldots+c_1 p_1+\ldots$, truncated to
a certain degree. One determines then the conditions on the
parameters $a_i, b_i,\ldots$ in order that $\widehat V(p)$ be
bounded below in each of the strata of ${\cal S}$, determines
these constrained minima, and compares them to single out the
absolute minimum. At the end one knows, for any set of values for
the parameters, which is the stable phase, that is the stratum
hosting the absolute minimum of $\widehat V(p)$. It is evident
that polynomial potentials of small degrees cannot develop minima
in certain strata of ${\cal S}$ for any set of values of the
parameters.

The order parameters are linked to some physical observables. This
is the case, for instance, of the number of pairs
$n=|\eta_{+}|^2+|\eta_{-}|^2=p_3$, of the number
$z=|\eta_{+}|^2-|\eta_{-}|^2=\pm \sqrt{p_3^2-p_2}$ that changes
sign under time reversal and determines the magnetic properties of
the crystal,~\cite{toltol} and of the components of the
deformation tensor of the unit cell of the crystal. From
Table~\ref{tab1} one sees that in the strata $\sigma_0$,
$\sigma_{1-}$, $\sigma_{1+}$, $\sigma_2$, and only in them, $z$
vanishes identically, so these strata cannot exhibit magnetic
properties. The change of the sign of $p_1$ does not modify the
\wPm\ and the orbit space ${\cal S}$, but changes the strata
$\sigma_{1-}\leftrightarrow \sigma_{1+}$ and
$\sigma_{2-}\leftrightarrow \sigma_{2+}$. As this sign change has
no physical meaning, the strata $\sigma_{1-}$ and $\sigma_{2-}$
are not physically distinguishable from $\sigma_{1+}$ and
$\sigma_{2+}$, nevertheless they are distinct strata lying in
different positions of the orbit space, and phase transitions
concerning each one of these strata are possible.

The \wPm\ and the orbit space here described, also appears in the
case of superconducting cubic crystals, if the order parameters
are assumed to transform according to the two-dimensional complex
representation of the cubic symmetry group.~\cite{guf96} In that
case however the isotropy groups are no longer those reported in
Table~\ref{tab2}. As already noted, many different symmetry groups
may give rise to the same \wPm\ and the same orbit space. When
this happens, the number of different symmetry phases and the list
of the phase transitions allowed for these systems are the same
(but the corresponding isotropy groups are generally different,
depending on the specific high symmetry group considered).
Moreover, for a given form of the potential in terms of the MIB,
the conditions to realize the listed phase transitions impose the
same conditions on the parameters of the potential.


\begin{thebibliography}{99}

\bibitem{guf82} Yu.M. Gufan, {\em Structural Phase Transitions}, (Nauka,
Moskow, 1982) (in Russian).

\bibitem{toltol} J.C. Tol\'edano and P. Tol\'edano, {\em The Landau Theory
of Phase Transitions}, (World Scientific, Singapore, 1987).

\bibitem{kosc} J. Koci\'nski, {\em Commensurate and Incommensurate Phase
Transitions}, (PWN, Warszawa, 1990).

\bibitem{vol-gor} G.E. Volovik and L.P. Gor'kov, {\em Sov. Phys.
JETP} {\bf 61}, 843 (1985).

\bibitem{sgr-ueda} M. Sgrist and K. Ueda, {\em Rev. Mod. Phys.}
{\bf 63}, 239 (1991).

\bibitem{naim} M. Na\"\i mark, A. Stern {\em Th\'eorie des Repr\'esentations des
Groupes}, (MIR, Moscou, 1979).

\bibitem{hamer} M. Hamermesh {\it Group Theory and its Application to
 Physical Problems}, (Addison-Wesley, Reading, MA, 1962).

\bibitem{as2} M. Abud and G. Sartori, {\it Ann. Phys.} {\bf 150}, 307 (1983).

\bibitem{st-cmp} G. Sartori and V. Talamini, {\it Commun. Math. Phys.} {\bf 139},
559 (1991).

\bibitem{guf71} Yu. M. Gufan, {\it Sov. Phys. - Solid State} {\bf 13}, 175 (1971).

\bibitem{guf94} Yu.M. Gufan, {\it Cryst. Rep.} {\bf 39}, 337 (1994).

\bibitem{bred} G.E. Bredon, {\em Introduction to Compact Transformation
Groups}, (Academic Press, New York, 1972).

\bibitem{mostow} G.D. Mostow, {\it Ann. Math.} {\bf 65}, 432 (1957).

\bibitem{as1} M. Abud and G. Sartori, {\it Phys. Lett.} B {\bf 104}, 147 (1981).

\bibitem{hilb} D. Hilbert, {\it Math. Ann.} {\bf 36}, 473 (1890).

\bibitem{weyl} H. Weyl, {\em The Classical Groups, their Invariants
and Representations}, (Princeton Univ. Press, Princeton, N. J.,
1939).

\bibitem{bierst-jdg} E. Bierstone, {\it J. Differential Geometry} {\bf 10}, 523 (1975).

\bibitem{schw-top} G.W. Schwarz, {\it Topology} {\bf 14}, 63 (1975).

\bibitem{mont-yang} D. Montgomery and C.T. Yang, {\it Trans. Am. Math. Soc.} {\bf 87}, 284 (1958).

\bibitem{bierst-top} E. Bierstone, {\it Topology} {\bf 14}, 245 (1975).

\bibitem{ps} C. Procesi and G.W. Schwarz, {\it Invent. Math.} {\bf 81}, 539 (1985).

\bibitem{sar-mpl} G. Sartori, {\it Mod. Phys. Lett.} A {\bf 4}, 91 (1989).

\bibitem{st-jgtp} G. Sartori and V. Talamini, {\it J. of Group Theory in Physics}
{\bf 2}, 13 (1994), avalaible at {\tt arXiv:hep-th/9512067}.

\bibitem{st-jmp} G. Sartori and V. Talamini, {\it J. Math. Phys.} {\bf 39}, 2367 (1998).

\bibitem{guf96} Yu.M. Gufan, {\it Cryst. Rep.} {\bf 41}, 381 (1996).

\end{thebibliography}
\end{document}